\documentclass[12pt]{article}
\usepackage{latexsym}
\usepackage{amsmath,,calrsfs}
\usepackage{amsfonts}
\usepackage{amssymb}
\usepackage{amscd}
\usepackage{bbm}
\usepackage{fancybox}
\usepackage{cite}
\usepackage{amsmath,amsfonts,amsbsy}
\usepackage{pstricks,pst-node}
\usepackage[small,bf,hang]{caption2}
\usepackage{graphicx}
\usepackage{epsfig}
\usepackage{psfrag}
\usepackage{comment}

\usepackage{float}

\psset{unit=1.3cm,linewidth=.5pt,radius=.2}  

\usepackage{multirow}                     
\usepackage{float}                          
\usepackage{lscape}                         
\usepackage{bm}

\newcommand{\beq}{\begin{equation}}
\newcommand{\eeq}{\end{equation}}
\newcommand{\bi}{\begin{itemize}}
\newcommand{\ei}{\end{itemize}}
\newcommand{\bt}{\begin{tabular}}
\newcommand{\et}{\end{tabular}}
\newcommand{\bc}{\begin{center}}
\newcommand{\ec}{\end{center}}

\newcommand{\be}{\begin{equation}}
\newcommand{\ee}{\end{equation}}
\newcommand{\bea}{\begin{eqnarray}}
\newcommand{\eea}{\end{eqnarray}}
\newcommand{\ba}{\begin{array}}
\newcommand{\ea}{\end{array}}

\def\bbox{{\,\lower0.9pt\vbox{\hrule \hbox{\vrule height 0.2 cm
\hskip 0.2 cm \vrule height 0.2 cm}\hrule}\,}}
\newcommand{\dsl}{\pa \kern-0.5em /}

\newcommand{\rmd}{\mathrm{d}}

\begin{document}

\begin{titlepage}
\begin{center}

\hfill UG-11-09\\ \hfill MIT-CTP-4217\\ \hfill DAMTP-2011-13\\

\vskip 1.5cm

{\Large \bf Modes of Log Gravity}

\vskip 1cm

{\bf Eric A.~Bergshoeff$^1$, Olaf Hohm$^2$, Jan Rosseel$^1$ and Paul K.~Townsend$^3$}

\vskip 25pt

{\em $^1$ \hskip -.1truecm Centre for Theoretical Physics,
University of Groningen, \\ Nijenborgh 4, 9747 AG Groningen, The
Netherlands \vskip 5pt }

{email: {\tt E.A.Bergshoeff@rug.nl, j.rosseel@rug.nl}} \\

\vskip 15pt

{\em $^2$ \hskip -.1truecm Center for Theoretical Physics,
Massachusetts Institute of Technology, \\ Cambridge, MA 02139, USA
\vskip 5pt }

{email: {\tt ohohm@mit.edu}} \\

\vskip 15pt

{\em $^3$ \hskip -.1truecm Department of Applied Mathematics and Theoretical Physics,\\ Centre for Mathematical Sciences, University of Cambridge,\\
Wilberforce Road, Cambridge, CB3 0WA, U.K. \vskip 5pt }

{email: {\tt P.K.Townsend@damtp.cam.ac.uk}} \\

\end{center}

\vskip 0.5cm

\begin{center} {\bf ABSTRACT}\\[3ex]
\end{center}
The physical modes of a recently proposed $D$-dimensional ``critical gravity'', linearized about its
anti-de Sitter vacuum,  are investigated.  All ``log mode'' solutions, which we categorize as
`spin 2' or  `Proca',  arise as limits of the massive spin 2 modes of the non-critical theory.
The linearized Einstein tensor of a spin 2 log mode is itself a `non-gauge' solution of the
linearized Einstein equations whereas the linearized Einstein tensor of a Proca mode
takes the form of a linearized  general coordinate transformation. Our results suggest the
existence of a holographically dual  logarithmic conformal field theory.

\end{titlepage}

\newpage
%

\section{Introduction}

When considered as a theory of interacting massless spin 2 particles in  a 4-dimensional Minkowski background, Einstein's theory of gravity is non-renormalizable.  It can be made renormalizable by the addition to the standard Einstein-Hilbert (EH) action of  curvature-squared terms, but the price is a loss of unitarity  \cite{stelle1}. There  are two exceptional cases. First, by adding a Ricci scalar squared term (with an appropriate sign) one gets  a theory  equivalent to a scalar coupled to gravity, which is unitary but not renormalizable. Second, by adding a Weyl-tensor squared term one gets a theory that is neither unitary nor renormalizable.  Renormalizability requires improved high-energy behaviour for both the  spin 0 and spin 2 projections of the graviton propagator, and this requires the presence of both Ricci-scalar squared and Weyl-squared terms \cite{stelle1}.

The situation is different in three spacetime dimensions in the sense that one gets a unitary theory of gravitons, albeit  massive ones,  by the addition to the standard EH action  of a particular curvature-squared term,  obtained by contracting the Einstein tensor with the Schouten tensor; this has been dubbed ``new massive gravity'' (NMG) \cite{Bergshoeff:2009hq}. The extension to  a ``cosmological NMG'' theory  introduces a new dimensionless parameter $\lambda$, and it  has been shown that  a unitary theory of massive gravitons in an AdS background is thus obtained for a certain range of $\lambda$ \cite{Bergshoeff:2009aq}. There are similarities here to topologically massive gravity \cite{Deser:1981wh}, which involves the addition to the EH term of a Lorentz-Chern-Simons term, and this may also be added to NMG to yield a ``general massive gravity'' model. However such parity-violating terms have no natural extension to  higher dimensions and so will not play a role here.

The properties of cosmological NMG are most easily understood by using a formulation in which the curvature-squared terms, of 4th order in derivatives,  are replaced by terms of at most 2nd-order by introducing a symmetric tensor auxiliary field\footnote{This formulation is also useful for other applications, see  \cite{Hohm:2010jc}.}. Linearizing about a maximally symmetric background one then finds a quadratic action for the metric perturbation and the auxiliary tensor field.  For generic values of $\lambda$  this action can be diagonalized to produce  the sum of a linearized EH term, which propagates no degrees of freedom in three dimensions,  and a  Fierz-Pauli (FP) action for a massive spin 2 mode. The  form of the FP mass term,  which is crucial for unitarity, is what requires the original curvature-squared term to be the contraction of the Einstein and Schouten tensors, and for a certain range of values of $\lambda$ the overall sign of the action is also what is required for unitarity.  The same analysis can be carried out in a higher spacetime dimension \cite{Nakasone:2009bn} but then the linearized EH term propagates a massless spin 2 mode, and either it or the massive spin-2 mode (depending on the overall sign of the action) is a ghost.
This is why the construction of NMG only yields a unitary theory  in three dimensions.

 However, there is another feature of NMG, which works for {\sl arbitrary}  spacetime dimension. It turns out that for a critical\,\footnote{A similar critical value was found earlier for cosmological topological massive gravity \cite{Li:2008dq}.} value of $\lambda$, at the boundary of the unitarity region, the linearized gravitational field  becomes a Lagrange multiplier imposing a constraint on the
 linearized auxiliary field. This constraint implies (in three dimensions) that the linearized auxiliary field takes the form of a field dependent general coordinate transformation. This does not mean, however, that this field can be gauged away and,
indeed, it corresponds to an additional mode, the so-called logarithmic
mode. In higher dimensions the constraint becomes a dynamical equation that allows for a wider class
of solutions, which we analyze here. At the critical point, the massive modes of the non-critical theory  coincide with the massless modes, and new logarithmic modes appear to replace them.  In three dimensions, these logarithmic modes were discussed in e.g. \cite{Grumiller:2008qz} and their existence  led to the conjecture that three-dimensional critical gravity theories are dual to two-dimensional  logarithmic CFT's (see e.g. \cite{Grumiller:2008qz,Grumiller:2009sn,Grumiller:2010tj,Skenderis:2009nt,Alishahiha:2010bw}).

 Logarithmic solutions in the context of  the higher-dimensional critical gravity models were recently found  in  \cite{Alishahiha:2011yb,Gullu:2011sj}.  In  this paper, we study the logarithmic modes  in more detail.  We show that  they are of two types,  which we dub  `spin 2' and `Proca'  log modes. The number of  independent spin 2 log modes is given by  the number of polarization states of a {\it massless}  spin 2 field, while the  number of independent  Proca log modes is given by the number of polarization states of  a {\it massive}  spin 1 field. We  present
 explicitly  the logarithmic solutions of the  linearized  $D=4$ critical gravity. We will show that these log modes have  properties similar to
 those of the three-dimensional log modes that were crucial
 in conjecturing the logarithmic CFT duals of three-dimensional critical gravities.

\section{The Model}

We consider the  following $D$-dimensional gravity theory

\begin{equation} \label{actschouten}
S = \frac{1}{\kappa^2} \int \rmd^D x \sqrt{-g} \left[ \sigma R - 2 \lambda m^2 + \frac{1}{m^2} G^{\mu \nu} S_{\mu \nu}
+ \frac{1}{m^{\prime 2}}{\cal L}_{\text{GB}}\right] \,,
\end{equation}
where
\begin{equation}
{\cal L}_{\text{GB}} \equiv \left(R^{\mu\nu\rho\sigma}R_{\mu\nu\rho\sigma}
-4 R^{\mu\nu}R_{\mu\nu} + R^2\right)\, ,
\end{equation}
which is  the Gauss-Bonnet combination. The parameter
$\sigma=0,\pm 1$ is a dimensionless constant, $\lambda$ is a
dimensionless cosmological parameter and $m^2, m^{\prime 2}$ are arbitrary
parameters of dimension mass squared that may be positive  or
negative. Furthermore,  $G_{\mu\nu}$ is the Einstein tensor and
$S_{\mu\nu}$ is the $D$-dimensional Schouten tensor:
\begin{equation}
S_{\mu \nu} = \frac{1}{D-2} \left(R_{\mu \nu} - \frac{1}{2(D-1)} R\, g_{\mu \nu} \right) \,.
\end{equation}
The reason that we have allowed, starting from higher than four dimensions,
for the Gauss-Bonnet \cite{Lovelock:1971yv} term
${\cal L}_{\text{GB}}$ in \eqref{actschouten} is that the linearization of this term around a maximally symmetric background
only affects the coefficient of the Einstein term in the quadratic action (see eq.~\eqref{linresult} below) but does not lead to new 4th-order higher-derivative terms.

For $D=3$ the term ${\cal L}_{\text{GB}}$  vanishes identically and
the action \eqref{actschouten} is that of cosmological NMG
\cite{Bergshoeff:2009hq}.\,\footnote{The special case of
$D=3,\lambda=\sigma=0$ was discussed in \cite{Deser:2009hb}.} For
$D=4$ the term ${\cal L}_{\text{GB}}$ reduces to a total derivative. At this point it is
convenient to use the identity \cite{Bergshoeff:2010ad}

\begin{equation}
R^{\mu\nu\rho\sigma}R_{\mu\nu\rho\sigma}
-4 R^{\mu\nu}R_{\mu\nu} + R^2 = W^{\mu\nu\rho\sigma} W_{\mu\nu\rho\sigma} - 4(D-3)G^{\mu\nu}S_{\mu\nu}\,,
\end{equation}
where $W_{\mu\nu\rho\sigma}$ is the Weyl tensor. This identity is
valid for any $D\ge 3$, although both sides vanish identically for
$D = 3$. For $D=4$ this identity shows that the Einstein tensor
times the Schouten tensor  equals the square of the Weyl tensor, up
to a total derivative, and the action \eqref{actschouten}, for
$\sigma=1$, reduces to the critical gravity theory considered in
\cite{Lu:2011zk}. For general $D\ge 5$  and $\sigma =1$ the same
action reduces to the 2-parameter family of theories recently
considered in \cite{Deser:2011xc}.

To discuss the quadratic approximation to the action \eqref{actschouten} it is convenient to lower the number of  derivatives in the action.
For the $G^{\mu\nu}S_{\mu\nu}$ term  this is achieved by
introducing an auxiliary field $f_{\mu \nu}$ that is a symmetric
two-tensor \cite{Bergshoeff:2009hq}. For the Gauss-Bonnet combination a similar trick does not work, at least not with a two-tensor auxiliary field,
but it is also not needed here since, as we already mentioned above, this term does not lead to higher-derivative terms  in the
 quadratic action. In terms of $f_{\mu\nu}$  the action \eqref{actschouten} reads

\begin{eqnarray}\label{auxiliary}
S & = & \frac{1}{\kappa^2} \int \rmd^D x \sqrt{-g} \Big[\sigma R - 2 \lambda m^2 + \frac{1}{(D-2)} f^{\mu \nu} G_{\mu \nu} - \frac{m^2}{4(D-2)} \left(f^{\mu \nu} f_{\mu \nu} - f^2 \right) \nonumber \\ & & \qquad \qquad + \frac{1}{m^{\prime 2}}{\cal L}_{\text{GB}} \Big] \,.
\end{eqnarray}
Elimination of the auxiliary field leads one back to the original formulation (\ref{actschouten}).

We now consider the linearization of the theory defined by \eqref{auxiliary} around a maximally symmetric vacuum with background metric $\bar {g}_{\mu\nu}$ and cosmological constant $\Lambda$. For such a background the Ricci tensor, Ricci scalar and Einstein tensor are given by
\begin{equation}
\bar{R}_{\mu \nu} = \frac{2 \Lambda}{(D-2)} \bar{g}_{\mu \nu} \,, \qquad \bar{R} = \frac{2 D \Lambda}{(D-2)} \,, \qquad \bar{G}_{\mu \nu} = - \Lambda \bar{g}_{\mu \nu} \,.
\end{equation}
In general the cosmological constant $\Lambda$ is not equal to the  parameter $\lambda$ \cite{Boulware:1985wk}. The two are  related by
the relation
\begin{equation} \label{Ll}
   \frac{(D-4)}{(D-1)(D-2)}\left[\frac{1}{2m^2}  -\frac{2}{m^{\prime 2}}
(D-3)\right]\Lambda^2 -\Lambda \sigma + \lambda m^2= 0 \,.
\end{equation}
This is a quadratic equation in $\Lambda$ which, for given values of the parameters, has 0,1 or 2 solutions, except for $D=4$ and $\sigma\ne 0$  where $\Lambda$ is uniquely fixed.

We next expand the metric $g_{\mu \nu}$ and the auxiliary field $f_{\mu \nu}$ around their background values:
\begin{eqnarray} \label{linansatz}
g_{\mu \nu} & = & \bar{g}_{\mu \nu} + \kappa h_{\mu \nu} \nonumber \,, \\
f_{\mu \nu} & = & \frac{2}{m^2 (D-1)} \left[ \Lambda (\bar{g}_{\mu \nu} + \kappa h_{\mu \nu}) + \kappa k_{\mu \nu} \right] \,.
\end{eqnarray}
The linearized Ricci tensor is given by

\begin{equation}
R_{\mu \nu} = \bar{R}_{\mu \nu} + \kappa R^{(1)}_{\mu \nu} + \kappa^2 R^{(2)}_{\mu \nu} + \mathcal{O}(\kappa^3) \,,
\end{equation}
where

\begin{equation}
R^{(1)}_{\mu \nu} = -\frac{1}{2} (\Box h_{\mu \nu} - \nabla^\rho \nabla_\mu h_{\rho \nu} - \nabla^\rho \nabla_\nu h_{\rho \mu} + \nabla_\mu \nabla_\nu h) \,,
\end{equation}
and

\begin{equation}
\bar{g}^{\mu \nu} R^{(2)}_{\mu \nu} = \frac{1}{2}h^{\mu \nu} \left(R^{(1)}_{\mu \nu} - \frac{1}{2} \bar{g}_{\mu \nu} \bar{g}^{\rho \sigma} R^{(1)}_{\rho \sigma} \right) + \mathrm{total} \ \mathrm{derivatives} \,.
\end{equation}
Linearizing \eqref{auxiliary} one finds that the terms linear in $1/\kappa$ cancel as a consequence of \eqref{Ll}.
The quadratic $\kappa$-independent terms lead to the following linearized Lagrangian

\begin{equation} \label{linresult}
\mathcal{L}_2   =   -\frac{1}{2} \bar\sigma
 h^{\mu \nu} \mathcal{G}_{\mu \nu}(h)
  + \frac{2}{m^2 (D-1)(D-2)} k^{\mu \nu} \mathcal{G}_{\mu \nu}(h)   - \frac{1}{m^2 (D-2)(D-1)^2} (k^{\mu \nu}k_{\mu \nu} - k^2) \,,
\end{equation}
where

\begin{equation}\label{barsigma}
\bar\sigma(\Lambda) \equiv \sigma -\frac{\Lambda}{m^2}\frac{1}{D-1}+4\frac{\Lambda}{m^{\prime 2}}\frac{(D-3)(D-4)}{(D-1)(D-2)}
\end{equation}
and where we have defined the Einstein operator

\begin{equation} \label{einsteinop}
\mathcal{G}_{\mu \nu}(h) = R^{(1)}_{\mu \nu} - \frac{1}{2} \bar{g}_{\mu \nu} \bar{g}^{\rho \sigma} R^{(1)}_{\rho \sigma} - \frac{2 \Lambda}{(D-2)} h_{\mu \nu} + \frac{\Lambda}{(D-2)} \bar{g}_{\mu \nu} h \,.
\end{equation}
The linearized Lagrangian \eqref{linresult} is invariant under the linearized diffeomorphisms

\begin{equation}\label{gtr}
\delta h_{\mu \nu} = \nabla_\mu \epsilon_\nu + \nabla_\nu \epsilon_\mu \,.
\end{equation}
This may be verified using the relation

\begin{equation}
[\nabla_\mu,\nabla_\nu] V_\rho = \frac{2\Lambda}{(D-1)(D-2)}\left ( {\bar g}_{\mu\rho}V_\nu - {\bar g}_{\nu\rho}V_\mu\right)
\end{equation}
for any vector $V_\mu$. The expansion of $f_{\mu \nu}$ in
\eqref{linansatz} is defined such that $k_{\mu \nu}$ is  gauge
invariant. Note also that the Einstein operator
\eqref{einsteinop} is gauge invariant. For $D=3$ the above result
agrees with the one given in \cite{Bergshoeff:2009aq}.

For general values of the parameters the first term in
\eqref{linresult} corresponds to a linearized Einstein-Hilbert term,
the second term provides a coupling between the $k$- and
$h$-fluctuation, while the last term provides a Fierz-Pauli mass
term for the $k$-fluctuation. After a diagonalization of the second
term, one deduces that the theory describes one massless graviton,
described by the linearized Einstein term  and one massive graviton,
described by a Fierz-Pauli Lagrangian \cite{Fierz-Pauli} with mass
given by

\begin{equation}\label{massivegraviton}
M^2 = -m^2(D-2)\bar\sigma\,.
\end{equation}
The kinetic terms of the massless and massive gravitons have opposite signs and therefore the theory is plagued with ghosts.

Following \cite{Lu:2011zk,Deser:2011xc} we now observe that at the critical point defined by the following special value of the cosmological
constant

\begin{equation}\label{crit}
\bar \sigma (\Lambda_{\text{crit}})=0
 \end{equation}
 the first term in \eqref{linresult}, i.e.~the linearized Einstein-Hilbert term, drops out.
The resulting critical Lagrangian is given by

\begin{equation}\label{criticalL}
m^2(D-1)(D-2)\,{\cal L}_{\text{crit}} =2\, h^{\mu\nu}{\cal G}_{\mu\nu}(k)-\frac{1}{(D-1)}\left(k^{\mu\nu}k_{\mu\nu}-k^2\right)\,.
\end{equation}
The field equation for  $h_{\mu \nu}$ is therefore

\begin{equation} \label{constraintk}
\mathcal{G}_{\mu \nu}(k) = 0 \,,
\end{equation}
while the $k$-equation of motion reads

\begin{equation}\label{eomk}
\mathcal{G}_{\mu \nu}(h) - \frac{1}{(D-1)} \left(k_{\mu \nu} -
\bar{g}_{\mu \nu} k \right) = 0 \,.
\end{equation}
By acting on \eqref{eomk} with $\nabla^\mu$ and using the
Bianchi identity $\nabla^\mu \mathcal{G}_{\mu \nu}(h) = 0$, we find

\begin{equation} \label{condk}
\nabla^\mu k_{\mu \nu} - \nabla_\nu k = 0 \,.
\end{equation}
Next, by taking the trace of \eqref{constraintk} and using
\eqref{condk} one finds that $\Lambda_{\text{crit}}\, k = 0$ and hence that

\begin{equation}\label{k=0}
k=0\;,
\end{equation}
provided that $\Lambda_{\text{crit}}\ne 0$, which we will assume to be the case from now on.
Substituting $k=0$  into  eq.~\eqref{eomk} it follows that

\begin{equation}\label{solutionk}
k_{\mu \nu} = (D-1)
\mathcal{G}_{\mu \nu}(h)\,.
\end{equation}
Finally, substituting \eqref{solutionk} into \eqref{constraintk}, one
finds that $h$ obeys the following fourth order equation

\begin{equation}\label{eom}
\mathcal{G}_{\mu \nu}(\mathcal{G}(h)) = 0 \,,
\end{equation}
together with the constraint

\begin{equation}\label{constrainth}
{\bar g}^{\mu\nu}{\cal G}_{\mu\nu}(h) = 0\,.
\end{equation}

The equations of motion \eqref{eom} which state that the Einstein operator of the Einstein operator of $h_{\mu\nu}$ is zero,
can be further simplified by imposing the gauge condition

\begin{equation} \label{gaugeh}
\nabla^\mu h_{\mu \nu} - \nabla_\nu h = 0 \,.
\end{equation}
Substituting this gauge condition
into the constraint \eqref{constrainth} one finds that
\begin{equation}
\Lambda_{\text{crit}}\, h  = 0\,.
\end{equation}
Since we already assumed that $\Lambda_{\text{crit}}\ne0$ we deduce that $h=0$ and hence we find that
\begin{equation} \label{simplcond}
\nabla^\mu h_{\mu \nu} = h = 0\,.
\end{equation}
Using this,  the linearized Einstein tensor reduces to

\begin{equation} \label{lineinst}
{\cal G}_{\mu\nu}(h) = -\frac{1}{2}\left ( \Box -\frac{4\Lambda}{(D-1)(D-2)}\right )h_{\mu\nu}\,.
\end{equation}
It follows  that the 4th-order operator appearing in the equation of motion \eqref{eom}
factorizes into two identical second order operators \cite{Lu:2011zk,Deser:2011xc}
\begin{equation}\label{final}
\left(\Box - \frac{4 \Lambda}{(D-1)(D-2)} \right)^2 h_{\mu \nu} = 0\,.
\end{equation}

\section{Massive, Massless and Log Modes}

We wish to analyze the solutions to the equations of motion
\eqref{final} assuming that we have an $\text{AdS}_D$ vacuum
solution with $\Lambda_{\text{crit}}<0$ and isometry algebra
$\text{SO}(2,D-1)$. 
The case of $D=4$ is of particular interest because of the improved
short-distance behavior of curvature-squared theories in this dimension,
as discussed in the introduction. We therefore focus on this case,
for which we infer from (\ref{barsigma}) that
\begin{equation}
\Lambda_{\text{crit}}(D=4) \ = \ 3m^2\sigma\,.
\end{equation}
The generalization to $D>4$ will be apparent while the $D=3$ case has been discussed in \cite{Bergshoeff:2009aq}.
Since away from the critical point there are both massless and massive
gravitons it is convenient to consider both types of solutions since this will facilitate the construction of the so-called log solutions
at the critical point.
To determine the massive solutions we follow the presentation of \cite{Li:2008dq}. Next, the massless modes are obtained by taking the massless limit of the massive ones and the log modes, which are valid solutions at the critical point only,  are obtained
by applying the limiting procedure of \cite{Grumiller:2008qz}. These log modes are solutions to the equations of motion \eqref{final} that are not annihilated by the separate second order Einstein operators.

In general, we expect for $D>3$ three classes of solutions at the critical point.
\bigskip

\noindent (1)\ \ The first class of solutions are the massless  gravitons,  which correspond to solutions of the  homogeneous equation

\begin{equation}\label{hom}
k_{\mu\nu}=3\,{\cal G}_{\mu\nu}(h)=0\,.
\end{equation}
\vskip .3truecm

\noindent (2)\ \ The second class of solutions will be called Proca
log modes and solve the inhomogeneous equation

\begin{equation}\label{inhom1}
k_{\mu\nu} = 3 {\cal G}_{\mu\nu}(h) = 2 \nabla_{(\mu}A_{\nu)}\,,
\end{equation}
for some vector field $A_\mu$. Written in terms of $k_{\mu\nu}$ they
are solutions of the massless Einstein equations ${\cal
G}_{\mu\nu}(k)=0$ that take the form of a field-dependent general
coordinate transformation.\,\footnote{This does not mean that $k_{\mu\nu}$ can be gauged away since
$k_{\mu\nu}$ is gauge invariant.} 
By substituting
\eqref{inhom1} into \eqref{condk}, one finds that $A_\mu$ satisfies the  equations of motion that follow from the following Proca Lagrangian
\cite{Bergshoeff:2009aq}:

\begin{equation}\label{Lcrit}
{\cal L}_{\text{Proca}} = -\frac{1}{4m^2}F^{\mu\nu}F_{\mu\nu} +
3\sigma A^\mu A_\mu\,, \hskip 1truecm F_{\mu\nu}=2
\partial_{[\mu}A_{\nu]}\,,
\end{equation}
which is why we dubbed the corresponding modes Proca modes.

\noindent (3)\ \ The third class of solutions will be denoted as
spin 2 log modes and correspond to solutions of the inhomogeneous
equation

\begin{equation} \label{thirdclass}
{\cal G}_{\mu\nu}(h) = k_{\mu\nu}^\perp\,,\hskip 2truecm k_{\mu\nu}^\perp \ne \nabla_{(\mu}A_{\nu)}\,.
 \end{equation}
In terms of $k_{\mu\nu}^\perp$ they correspond to non-trivial solutions of the massless Einstein equations ${\cal G}_{\mu\nu}(k^\perp)=0$. Strictly speaking, eq.~\eqref{thirdclass} defines an equivalence class of solutions
since to every spin 2 log mode one can add a Proca mode.

\bigskip

We now study the solutions of the linearized equations of motion {\it away from the critical point},
following  the group theoretical approach  of \cite{Li:2008dq}.   Our starting point is
 the $\mathrm{AdS}_4$ metric which in global coordinates $(\tau\,,\rho\,,\theta\,,\phi)$ is given by\,:
\begin{equation}
\rmd s^2 = L^2 \left(-\rmd\tau^2 \cosh(\rho)^2 + \rmd \rho^2 +\sinh(\rho)^2\left(\rmd \theta^2 +\rmd\phi^2 \sin(\theta)^2\right)\right) \,.
\end{equation}
Here $L$ is related to  the cosmological constant $\Lambda$  by
\begin{equation}
\Lambda = -\frac{3}{L^2} \,.
\end{equation}
The isometry group of $\mathrm{AdS}_4$ is given by $\mathrm{SO}(2,3)$ which is generated by 10 Killing vectors. These Killing vectors can be grouped into Cartan generators and positive and negative root generators of $\mathrm{SO}(2,3)$. The  Killing vectors corresponding to the two Cartan generators are given by:
\begin{equation}
H_1 = i \partial_\tau \,, \qquad H_2 = -i \partial_\phi \,.
\end{equation}
The  Killing vectors corresponding to the four positive roots will be taken to be
\begin{eqnarray}
E^{\alpha_1} & = & \frac{1}{2} e^{i (\tau +\phi )} \sin(\theta) \tanh(\rho) \partial_\tau - \frac{1}{2} i e^{i (\tau +\phi )} \sin(\theta) \partial_\rho \nonumber \\ & & -  \frac{1}{2} i e^{i (\tau +\phi )} \cos(\theta) \coth(\rho) \partial_\theta  + \frac{1}{2} e^{i (\tau +\phi )} \coth(\rho) \csc(\theta) \partial_\phi\,, \nonumber \\ E^{\alpha_2} & = & -i e^{i \phi } \partial_\theta + e^{i \phi } \cot(\theta) \partial_\phi\,, \nonumber \\ E^{\alpha_3} & = & \frac{1}{2} e^{i (\tau -\phi )} \sin(\theta) \tanh(\rho) \partial_\tau  - \frac{1}{2} i e^{i (\tau -\phi )} \sin(\theta) \partial_\rho\,, \nonumber \\Ê& & - \frac{1}{2} i e^{i (\tau -\phi )} \cos(\theta) \coth(\rho) \partial_\theta -\frac{1}{2} e^{i (\tau -\phi )} \coth(\rho) \csc(\theta) \partial_\phi \nonumber\,, \\ E^{\alpha_4} & = & e^{i \tau } \cos(\theta) \tanh(\rho) \partial_\tau - i e^{i \tau } \cos(\theta) \partial_\rho + i e^{i \tau } \coth(\rho) \sin(\theta) \partial_\theta \,.
\end{eqnarray}
The Killing vectors corresponding to the four negative roots are proportional to  the complex conjugates of the above four Killing vectors:
\begin{equation}
E^{-\alpha_1} =  (E^{\alpha_1})^* \,, \quad E^{-\alpha_2} = -(E^{\alpha_2})^* \,, \quad E^{-\alpha_3} = (E^{\alpha_3})^* \,, \quad E^{-\alpha_4} =  (E^{\alpha_4})^* \,.
\end{equation}
The root vectors corresponding to the above positive roots are given by
\begin{equation}
\alpha_1 = (-1,1)\,, \quad \alpha_2 = (0,1) \,, \quad \alpha_3 = (-1,-1) \,, \quad \alpha_4 = (-1,0) \,.
\end{equation}
The above Killing vectors are normalized in the Cartan-Weyl fashion, i.e.~the following commutation relations hold
\begin{eqnarray}
& & [H_i, H_j] =  0 \,, \hskip 3truecm  i = 1,2 \,,\nonumber \\ & & [ H_i, E^{\alpha_x}] = \alpha_x^i E^{\alpha_x} \,,  \hskip 1.8truecm x = 1, \cdots, 4 \,,  \nonumber \\ & & [E^{\alpha_x},E^{-\alpha_x}]  =  \frac{2}{|\alpha_x|^2} \alpha_x \cdot H \,,
\end{eqnarray}
where
\begin{equation}
|\alpha_x|^2 = \sum_{i=1}^2 (\alpha_x^i)^2 \,.
\end{equation}
The Casimir operator $\mathcal{C}$ can then be constructed as follows
\begin{equation}
\mathcal{C} = \sum_{i=1}^2 H_i H_i + \sum_{x=1}^4 \frac{|\alpha_x|^2}{2} (E^{\alpha_x} E^{-\alpha_x} + E^{-\alpha_x} E^{\alpha_x}) \,.
\end{equation}
When acting on a scalar field $S(\tau,\rho,\theta,\phi)$ the Casimir operator is given by
\begin{equation}
\mathcal{C} S = L^2 \nabla^2 S \,.
\end{equation}
Similarly, when acting  on a metric perturbation $h_{\mu\nu}$ the Casimir operator is given by
\begin{equation}
\left(\mathcal{C} - 8  \right) h_{\mu \nu} = L^2\nabla^2 h_{\mu \nu}\,.
\end{equation}

We now consider the $D=4$ linearized equations of motion away from
the critical point
\begin{equation}
\left(\nabla^2 + \frac{2}{L^2} - M^2 \right) \left(\nabla^2 + \frac{2}{L^2} \right) h_{\mu \nu} = 0 \,,
\end{equation}
where $M^2 = -m^2\bar \sigma$ is the mass of the graviton, see eq.~\eqref{massivegraviton}.
These equations can be rewritten in terms of the Casimir operator as follows:
\begin{equation} \label{eomC}
(\mathcal{C} - 6 - L^2 M^2) (\mathcal{C} - 6) h_{\mu \nu} = 0 \,.
\end{equation}
We now look for a metric perturbation $\psi_{\mu \nu}$ that forms a highest weight state, with $M^2\ne 0$, of the $\mathrm{SO}(2,3)$ isometry algebra. This state is an eigenstate of $H_1$ and $H_2$ (acting as Lie derivatives) with eigenvalues $E_0$ and $s$:
\begin{equation}
H_1 \psi_{\mu \nu} = E_0 \psi_{\mu \nu} \,, \quad H_2 \psi_{\mu \nu} = s \psi_{\mu \nu} \,,
\end{equation}
while it is annihilated by all positive roots $E^{\alpha_x}$
\begin{equation}
E^{\alpha_x} \psi_{\mu \nu} = 0 \,, \qquad x = 1, \cdots, 4 \,.
\end{equation}
Using the conditions \eqref{simplcond}, i.e.
\begin{equation}
\nabla^\mu \psi_{\mu \nu} = 0 \,, \qquad \bar{g}^{\mu \nu} \psi_{\mu \nu} = 0 \,,
\end{equation}
we find that a solution for the highest weight state can be found for
\begin{equation}
s = 2 \,.
\end{equation}
The explicit expression of the solution reads\,\footnote{Similar AdS wave solutions for the full nonlinear theory have recently been considered in \cite{Gullu:2011sj}.}

\begin{eqnarray} \label{hwstate}
\psi_{\tau \tau} & = & \ \ - \psi_{\tau \phi}\ \  = \ \ \psi_{\phi \phi} \ \ = \ \ e^{-i E_0 \tau +2 i \phi } \sin(\theta)^2 \sinh(2 \rho)^{\frac{2-E_0}{2}} \tanh(\rho)^{1+\frac{E_0}{2}} \,, \nonumber \\ \psi_{\tau \rho} & = & \ \ -\psi_{\rho \phi} \ \ = \ \  i\, \mathrm{csch}(\rho) \, \mathrm{sech}(\rho)\, \psi_{\tau \tau} \,, \nonumber \\ \psi_{\tau \theta} & = & \ \ -\psi_{\theta \phi} \ \ = \ \ i\, \cot(\theta) \psi_{\tau \tau}  \\\psi_{\rho \rho} & = & -4\, \mathrm{csch}(2\rho)^2 \, \psi_{\tau \tau} \,, \nonumber \\ \psi_{\rho \theta} & = & -2 \cot(\theta) \mathrm{csch}(2 \rho) \psi_{\tau \tau} \,, \nonumber \\Ê\psi_{\theta \theta} & = & - \cot(\theta)^2
\psi_{\tau\tau} \,.\nonumber
\end{eqnarray}
Using that on a highest weight state
\begin{equation}
\mathcal{C} \psi_{\mu \nu} = \left( E_0 (E_0 - 3) + s(s+1) \right) \psi_{\mu \nu} \,,
\end{equation}
we find from the equation of motion (\ref{eomC}) that $E_0$ has to obey
\begin{equation}
\left( E_0 (E_0 - 3) - L^2 M^2 \right) E_0 (E_0 - 3) = 0 \,.
\end{equation}

The descendant states of (\ref{hwstate}) can be obtained by acting with Killing vectors corresponding to the negative roots. There is an infinite number of descendant states, but they can be organized  in representations of $\mathrm{SO}(3)$. Indeed, the negative root $E^{-\alpha_2}$ only lowers the $s$-eigenvalue, while it leaves $E_0$ untouched. $H_2$, $E^{\alpha_2}$ and $E^{-\alpha_2}$ thus form the algebra of the compact $\mathrm{SO}(3)$ subgroup of $\mathrm{SO}(2,3)$ and the descendant states organize themselves in representations of this $\mathrm{SO}(3)$ subgroup.
By acting with $E^{-\alpha_2}$ on (\ref{hwstate}) one thus obtains five solutions of the equations of motion (\ref{eomC}), that form a spin-2 $\mathrm{SO}(3)$ multiplet, with $s=+2$, $+1$, $0$, $-1$ and $-2$, respectively. In principle we can now determine all descendant solutions.
In practice,  it  is often enough to restrict to the highest weight state and the above $\text{SO}(3)$ descendants. This finishes our discussion of the massive solutions.

The massless solutions are obtained by taking the
limit $M \rightarrow 0$ of the massive ones. The resulting massless solutions solve the equations
\begin{equation} \label{eommassless}
\left(\nabla^2 + \frac{2}{L^2}\right) h_{\mu \nu} = 0 \,.
\end{equation}
In the massless limit we must have $E_0=0$ or $E_0=3$. From the
first line in \eqref{hwstate} we see that for $E_0 = 0$ the solution
blows up  for $\rho \rightarrow + \infty$, while for $E_0 = 3$ the
solution is well-behaved in this limit. In the following we will
mainly concentrate on solutions that fall off to zero in the $\rho
\rightarrow + \infty$ limit. In the massless limit, with  $E_0 =
3$, the five  massive solutions, with $s=+2\,,\cdots \,,s=-2$, all
become non-zero solutions of the Einstein equations
(\ref{eommassless}). This happens for the $s=+2$ and
$s=-2$ solutions,  in particular,  but also for the $s=+1$, $0$ and $-1$
solutions. As the massless Einstein equations describe only two
helicity-2 modes, it is to be expected that only 2 linear
combinations of the above modes correspond to physical modes,
belonging to the first class of solutions at the critical point
described in \eqref{hom}. Three other linear combinations are then
expected to correspond to infinitesimal general coordinate
transformations.

Having discussed the massive and massless modes we now  consider the
log modes. As in the three-dimensional case, one expects logarithmic
modes to show up that are solutions of the fourth order equation of
motion, but that do not solve (\ref{eommassless}). Applying the
limiting procedure of \cite{Grumiller:2008qz} on the highest weight
state, we find the following logarithmic mode:\,\footnote{An
alternative expression for such a log mode has recently been given
in \cite{Alishahiha:2011yb}.}
\begin{equation} \label{logstate}
\psi_{\mu\nu}^{\mathrm{log}}(s=2) = f(\tau,\rho)\, \psi^{(2)}_{\mu \nu}(E_0=3) \,,
\end{equation}
with

\begin{equation}\label{f}
f(\tau,\rho) = \frac{1}{2} (-2 i \tau -\log(\sinh(2 \rho))+\log(\tanh(\rho)))\
\end{equation}
and  where $\psi^{(2)}_{\mu \nu}(E_0=3)$ denotes the $s=2$ solution (\ref{hwstate}) taken at the massless point $E_0=3$. One can check that (\ref{logstate}) is traceless and has zero divergence. The Einstein tensor of this log mode can thus be calculated via (\ref{lineinst}) and we  find that it
 is given by
\begin{equation}
\mathcal{G}_{\mu \nu}(\psi^{\mathrm{log}}(s=2)) = -\frac{3}{2 L^2}
\psi^{(2)}_{\mu \nu}(E_0=3) \,.
\end{equation}

The above features of the $s=2$ log state persist for all five spin-2 states. In all cases the log mode solution is given by
\begin{equation}\label{logstates=1}
\psi_{\mu\nu}^{\mathrm{log}}(s) = f(\tau,\rho)\, \psi^{(s)}_{\mu \nu}(E_0=3) \,,
\end{equation}
with $f(\tau,\rho)$ given by \eqref{f} and where $\psi^{(s)}_{\mu
\nu}(E_0=3)$ is the helicity $s$ solution of the massless Einstein
equation \eqref{eommassless}. These five log modes form a 5-plet
under $\text{SO}(3)$ and are related to each other by the raising
and lowering operators of $\text{SO}(3)$. We have checked that in
all cases the Einstein tensor of $\psi_{\mu\nu}^{\mathrm{log}}(s)$ is proportional to the
helicity $s$ solution of the massless Einstein equation:
\begin{equation}
\mathcal{G}_{\mu \nu}(\psi^{\mathrm{log}}(s)) = -\frac{3}{2 L^2}
\psi^{(s)}_{\mu \nu}(E_0=3) \,.
\end{equation}
One thus expects that linear combinations of the log modes can be divided in two classes. Two linear combinations are such that their Einstein tensor gives rise to non-trivial solutions of
the Einstein equations. These are the so-called spin 2 log modes
that belong to the third class of solutions, defined in
(\ref{thirdclass}). Three linear combinations are then expected to
have an Einstein tensor that takes the form of an infinitesimal
general coordinate transformation. These three log modes are
therefore Proca log modes and belong to the second class of
solutions described in \eqref{inhom1}.

Finally, we mention some properties of the logarithmic modes that we have found. The
mode $\psi_{\mu\nu}^{\mathrm{log}}(s=2)$ is annihilated by all four
positive root generators:
\begin{equation}
E^{\alpha_x}\ \psi_{\mu\nu}^{\mathrm{log}}(s=2) = 0\,, \qquad
x=1,\cdots,4 \,.
\end{equation}
The other log modes in the 5-plet are obtained by acting with the
$\mathrm{SO}(3)$ lowering operator $E^{-\alpha_2}$. All log modes
correspond to eigenstates of $H_2$
\begin{equation} \label{h2beh}
H_2 \ \psi_{\mu\nu}^{\mathrm{log}}(s) = s\
\psi_{\mu\nu}^{\mathrm{log}}(s) \,,
\end{equation}
but they do not correspond to eigenstates of $H_1$
\begin{equation}\label{h1beh}
H_1 \ \psi_{\mu\nu}^{\mathrm{log}}(s) = 3
\psi_{\mu\nu}^{\mathrm{log}}(s) + \psi^{(s)}_{\mu \nu}(E_0=3) \,.
\end{equation}
This structure is reminiscent of the three-dimensional case
\cite{Grumiller:2008qz,Grumiller:2009sn,Grumiller:2010tj,Skenderis:2009nt}.
In that case, the analogue of the properties \eqref{h2beh},
\eqref{h1beh} led to the conjecture that the dual CFT is a
logarithmic one.

\section{Conclusions}

We have studied the recently proposed $D$-dimensional critical gravity theories of
\cite{Lu:2011zk,Deser:2011xc}. The family of models considered contains,
besides a dimensionless cosmological parameter $\lambda$ and two mass
parameters $m^2\,,m^{\prime 2}$, an additional dimensionless parameter $\sigma=0,\pm 1$ multiplying  the Einstein-Hilbert term. After linearization about a maximally-symmetric background, $\sigma$  is replaced by an effective EH coefficient  $\bar\sigma$. The critical theory is defined by  $\bar\sigma=0$, which condition determines the cosmological constant $\Lambda$, which we assumed to be negative.  The quadratic critical Lagrangian \eqref{criticalL} depends only on the mass parameter $m^2$. This allows for different choices of $\sigma$. In particular, one could take a ``wrong sign'' Einstein-Hilbert term in the starting action or even no Einstein-Hilbert term at all\footnote{Note, however, that a pure Weyl-squared term  is not possible, in view of \eqref{barsigma}.}.

At the critical point, the linearized equation of motion is essentially given by the Einstein tensor of the Einstein tensor of the metric perturbation; in others words, one acts twice on the perturbation with  the
``Einstein operator''  (defined by linearization of the Einstein tensor).  Any solution of the linearized Einstein equations  is therefore a solution, and these are the massless spin 2 modes. In addition, there are logarithmic solutions that are not annihilated by a single action of the  Einstein operator. We subdivided these logarithmic solutions into two classes: the  spin 2 and Proca  modes. For $D=4$, we  used the $\mathrm{SO}(2,3)$ isometry group of $\mathrm{AdS}_4$ to explicitly calculate the
massive and massless modes away from the critical point. We have shown  how, at the critical point, the massive modes are replaced by spin 2 and Proca log modes.

So far, our analysis has been done without a careful consideration
of the boundary conditions. As an example of how important boundary conditions can be, it is interesting to consider, for $D=3$, the relation between the Proca modes, propagated by the Lagrangian \eqref{Lcrit} and the Proca log modes. This relationship is not one-to-one. It turns out that the highest weight state of the Proca log mode corresponds to a non-normalizable solution of the equations of motion that follow from \eqref{Lcrit}. Its descendants, however, do give rise to normalizable solutions.\,\footnote{In the context of topological massive gravity such a descendant mode has been described in \cite{Giribet:2008bw}. Different formulations at the linearized level also occur in the case of topological massive gravity, see e.g.~the discussion in \cite{Carlip:2008jk,Carlip:2009ey}.}

The boundary conditions, when logarithmic modes are included,  have been
well-studied for the special case of $D=3$:  it has
been established \cite{Grumiller:2008qz} that the logarithmic bulk
modes require weaker boundary conditions than the Brown-Henneaux
ones \cite{Brown:1986nw}. These weaker boundary conditions were
dubbed `logarithmic boundary conditions' and they play an essential
role in the search for the two-dimensional CFT-duals of the various
three-dimensional massive gravities. The logarithmic modes for $D=4$ critical gravity
 studied here exhibit an  analogous group theoretical structure.
In the $D=3$  case, the existence and structure of these logarithmic modes lends support
for the conjecture that the CFT-dual of three-dimensional critical
gravity theories is of the logarithmic type (see e.g.
\cite{Grumiller:2009sn,Skenderis:2009nt,Grumiller:2010tj}).
It would be of interest to see whether one could similarly define a consistent set of
logarithmic boundary conditions in the higher-dimensional case and, if so, to see what one
could say about the CFT-duals of critical gravities in arbitrary dimensions.

\bigskip
\noindent{\bf Note added:}
Following submission of the original version of this paper to the arxiv,
a revised version of ref.~\cite{Lu:2011zk} appeared in which the log modes of 4D critical gravity
 presented here were found  to have positive energy (the massless Einstein modes
have zero energy). Although this is encouraging, it appears likely
that the log modes are not orthogonal to the Einstein modes, which
would imply the existence of linear combinations of negative norm,
as happens in critical TMG (see, e.g., sec.~4.1.2 of
\cite{Grumiller:2009mw}).\footnote{We thank Massimo Porrati for
pointing this out to us.} This would imply non-unitarity, as is to
be expected from the non-unitarity of the dual logarithmic CFT.

\bigskip
\bigskip


\noindent {\bf Acknowledgements}
\vskip .1truecm

We thank Diederik Roest and Massimo Porrati for helpful discussions and correspondence. The work of OH is supported by the
DFG-The German Science Foundation and in part by funds provided by the U.S.
Department of Energy (DoE) under the cooperative research agreement DE-FG02-
05ER41360.

\providecommand{\href}[2]{#2}\begingroup\raggedright\endgroup


\begin{thebibliography}{1}


\bibitem{stelle1} K.S. Stelle,
``Renormalization of higher-derivative quantum gravity'', Phys. Rev.
{\bf D16}, 953 (1977);
``Classical gravity with higher derivatives'', Gen. Rel. Grav. {\bf
9}, 353 (1978).


\bibitem{Bergshoeff:2009hq}
  E.~A.~Bergshoeff, O.~Hohm and P.~K.~Townsend,
  ``Massive Gravity in Three Dimensions,''
  Phys.\ Rev.\ Lett.\  {\bf 102} (2009) 201301
  [arXiv:0901.1766 [hep-th]].

\bibitem{Bergshoeff:2009aq}
  E.~A.~Bergshoeff, O.~Hohm and P.~K.~Townsend,
  ``More on Massive 3D Gravity,''
  Phys.\ Rev.\  D {\bf 79} (2009) 124042
  [arXiv:0905.1259 [hep-th]].


\bibitem{Deser:1981wh}
  S.~Deser, R.~Jackiw and S.~Templeton,
  ``Topologically massive gauge theories,''
  Annals Phys.\  {\bf 140} (1982) 372
  [Erratum-ibid.\  {\bf 185} (1988\ APNYA,185,406.1988\ APNYA,281,409-449.2000) 406.1988\ APNYA,185,406.1988].


\bibitem{Hohm:2010jc}
  O.~Hohm, E.~Tonni,
  ``A boundary stress tensor for higher-derivative gravity in AdS and Lifshitz backgrounds,''
  JHEP {\bf 1004}, 093 (2010).
  [arXiv:1001.3598 [hep-th]].


\bibitem{Nakasone:2009bn}
  M.~Nakasone, I.~Oda,
  ``On Unitarity of Massive Gravity in Three Dimensions,''
  Prog.\ Theor.\ Phys.\  {\bf 121 } (2009)  1389-1397.
  [arXiv:0902.3531 [hep-th]].



\bibitem{Li:2008dq}
  W.~Li, W.~Song and A.~Strominger,
  ``Chiral Gravity in Three Dimensions,''
  JHEP {\bf 0804} (2008) 082
  [arXiv:0801.4566 [hep-th]].

\bibitem{Grumiller:2008qz}
  D.~Grumiller and N.~Johansson,
  ``Instability in cosmological topologically massive gravity at the chiral
  point,''
  JHEP {\bf 0807} (2008) 134
  [arXiv:0805.2610 [hep-th]];
  S.~Ertl, D.~Grumiller and N.~Johansson,
  ``Erratum to `Instability in cosmological topologically massive gravity at
  the chiral point', arXiv:0805.2610,''
  arXiv:0910.1706 [hep-th].


\bibitem{Grumiller:2009sn}
  D.~Grumiller, O.~Hohm,
  ``AdS(3)/LCFT(2): Correlators in New Massive Gravity,''
  Phys.\ Lett.\  {\bf B686}, 264-267 (2010).
  [arXiv:0911.4274 [hep-th]].



\bibitem{Grumiller:2010tj}
  D.~Grumiller, N.~Johansson, T.~Zojer,
  ``Short-cut to new anomalies in gravity duals to logarithmic conformal field theories,''
  [arXiv:1010.4449 [hep-th]].

\bibitem{Skenderis:2009nt}
  K.~Skenderis, M.~Taylor, B.~C.~van Rees,
  ``Topologically Massive Gravity and the AdS/CFT Correspondence,''
  JHEP {\bf 0909}, 045 (2009).
  [arXiv:0906.4926 [hep-th]].


\bibitem{Alishahiha:2010bw}
  M.~Alishahiha, A.~Naseh,
  ``Holographic renormalization of new massive gravity,''
  Phys.\ Rev.\  {\bf D82 } (2010)  104043.
  [arXiv:1005.1544 [hep-th]].



\bibitem{Gullu:2011sj}
  I.~G\"ull\"u, M.~G\"urses, T.~C.~\c{S}i\c{s}man and B.~Tekin,
  ``AdS Waves as Exact Solutions to Quadratic Gravity,''
  [arXiv:1102.1921 [hep-th]].


\bibitem{Alishahiha:2011yb}
  M.~Alishahiha, R.~Fareghbal,
  ``D-Dimensional Log Gravity,''
   [arXiv:1101.5891 [hep-th]].

\bibitem{Lovelock:1971yv}
  D.~Lovelock,
  ``The Einstein tensor and its generalizations,''
  J.\ Math.\ Phys.\  {\bf 12 } (1971)  498-501.



\bibitem{Deser:2009hb}
  S.~Deser,
  ``Ghost-free, finite, fourth order D=3 (alas) gravity,''
  Phys.\ Rev.\ Lett.\  {\bf 103} (2009) 101302
  [arXiv:0904.4473 [hep-th]].



\bibitem{Bergshoeff:2010ad}
  E.~A.~Bergshoeff, O.~Hohm, P.~K.~Townsend,
  ``Gravitons in Flatland,''
[arXiv:1007.4561 [hep-th]].




\bibitem{Lu:2011zk}
  H.~Lu and C.~N.~Pope,
  ``Critical Gravity in Four Dimensions,''
  arXiv:1101.1971 [hep-th].


\bibitem{Deser:2011xc}
  S.~Deser, H.~Liu, H.~Lu, C.N.~Pope, T.\c{C}. \c{S}i\c{s}man and B. Tekin,
  ``Critical Points of D-Dimensional Extended Gravities,''
[arXiv:1101.4009 [hep-th]].

\bibitem{Boulware:1985wk}
  D.~G.~Boulware, S.~Deser,
  ``String Generated Gravity Models,''
  Phys.\ Rev.\ Lett.\  {\bf 55 } (1985)  2656.

\bibitem{Fierz-Pauli}
 M. Fierz and W. Pauli,
``On Relativistic Wave Equations for Particles of Arbitrary Spin in an Electromagnetic Field'',
Proc.  Roy. Soc. Series A, Vol. 173, No. 953, 211 (1939).



\bibitem{Giribet:2008bw}
  G.~Giribet, M.~Kleban and M.~Porrati,
  ``Topologically Massive Gravity at the Chiral Point is Not Chiral,''
  JHEP {\bf 0810} (2008) 045
  [arXiv:0807.4703 [hep-th]].


\bibitem{Carlip:2008jk}
  S.~Carlip, S.~Deser, A.~Waldron and D.~K.~Wise,
  ``Cosmological Topologically Massive Gravitons and Photons,''
  Class.\ Quant.\ Grav.\  {\bf 26} (2009) 075008
  [arXiv:0803.3998 [hep-th]].


\bibitem{Carlip:2009ey}
  S.~Carlip,
  ``Chiral Topologically Massive Gravity and Extremal B-F Scalars,''
  JHEP {\bf 0909} (2009) 083
  [arXiv:0906.2384 [hep-th]].





\bibitem{Brown:1986nw}
  J.~D.~Brown and M.~Henneaux,
  ``Central Charges in the Canonical Realization of Asymptotic Symmetries: An
  Example from Three-Dimensional Gravity,''
  Commun.\ Math.\ Phys.\  {\bf 104} (1986) 207.

\bibitem{Grumiller:2009mw}
  D.~Grumiller and I.~Sachs,
  ``AdS3/LCFT2 -- Correlators in Cosmological Topologically Massive
  Gravity,''
  JHEP {\bf 1003} (2010) 012
  [arXiv:0910.5241 [hep-th]].


\end{thebibliography}
\end{document}